\documentclass[letter,fleqn,usenatbib]{mnras}
\usepackage{savesym}
\usepackage{amsmath}
\savesymbol{iint}
\savesymbol{iiint}
\usepackage{txfonts}
\restoresymbol{TXF}{iint}
\restoresymbol{TXF}{iiint}

\usepackage[T1]{fontenc}
\usepackage{ae,aecompl}

\usepackage{graphicx}	% Including figure files
\usepackage{amsmath}	% Advanced maths commands
\usepackage{amssymb}	% Extra maths symbols
\usepackage{lscape}
\usepackage{booktabs}
\newcommand{\angstrom}{{\rm \AA}}

\newcommand{\HII}{H\,{\small II}}
\newcommand{\hbeta}{H{$\beta$}}
\newcommand{\halpha}{H{$\alpha$}}

\newcommand{\OIII}{[\ion{O}{III}]}
\newcommand{\loiii}{$L_{{\rm [\ion{O}{III}]}}$}
\newcommand{\OIIc}{[\ion{O}{II}]\,$\lambda\lambda$3726,3729}
\newcommand{\OIIIa}{[\ion{O}{III}]$\lambda$4959}
\newcommand{\OIIIb}{[\ion{O}{III}]$\lambda$5007}
\newcommand{\OIIIc}{[\ion{O}{III}]\,$\lambda\lambda$4959,5007}

\newcommand{\NIIb}{[\ion{N}{II}]\,$\lambda$6584}
\newcommand{\NeV}{[\ion{Ne}{V}]\,$\lambda$3426}
\newcommand{\NeIII}{[\ion{Ne}{III}]\,$\lambda$3869}

\title[Double-peaked Narrow Lines in Powerful AGNs]{A High Fraction of Double-peaked Narrow Emission Lines in Powerful Active Galactic Nuclei}

\author[LYU \& LIU]{
Yang Lyu,$^{1}$
Xin Liu,$^{2,3}$\thanks{E-mail: xinliuxl@illinois.edu} \\
$^{1}$Departments of Physics, University of Illinois, Urbana-Champaign, Urbana, IL 61820, USA\\
$^{2}$Department of Astronomy, University of Illinois at Urbana-Champaign, Urbana, IL 61801, USA\\
$^{3}$National Center for Supercomputing Applications, University of Illinois at Urbana-Champaign, 605 East Springfield Avenue, Champaign, IL 61820, USA
}

\date{Accepted 2016 August 02. Received 2016 July 22; in original form 2016 June 18}

\pubyear{2016}

\begin{document}
\label{firstpage}
\pagerange{\pageref{firstpage}--\pageref{lastpage}}
\maketitle

\begin{abstract}
One percent of redshift $z\sim0.1$ Active Galactic Nuclei (AGNs) show velocity splitting of a few hundred km s$^{-1}$ in the narrow emission lines in spatially integrated spectra. Such line profiles have been found to arise from the bulk motion of ionized gas clouds associated with galactic-scale outflows, merging pairs of galaxies each harboring a supermassive black hole (SMBH), and/or galactic-scale disk rotation. It remains unclear, however, how the frequency of narrow-line velocity splitting may depend on AGN luminosity. Here we study the correlation between the fraction of Type 2 AGNs with double-peaked narrow emission lines and AGN luminosity as indicated by \OIIIb\ emission-line luminosity \loiii . We combine the sample of \citet{Liu2010b} at $z\sim0.1$ with a new sample of 178 Type 2 AGNs with double-peaked \OIII\ emission lines at $z\sim0.5$. We select the new sample from a parent sample of 2089 Type 2 AGNs from the SDSS-III/Baryon Oscillation Spectroscopic Survey. We find a statistically significant ($\sim4.2\sigma$) correlation between \loiii\ and the fraction of objects that exhibit double-peaked narrow emission lines among all Type 2 AGNs, corrected for selection bias and incompleteness due to \OIII\ line width, equivalent width, splitting velocity, and/or equivalent width ratio between the two velocity components. Our result suggests that galactic-scale outflows and/or merging pairs of SMBHs are more prevalent in more powerful AGNs, although spatially resolved follow up observations are needed to resolve the origin(s) for the narrow-line velocity splitting for individual AGNs. 
\end{abstract}


%%%%%%%%%%%%%%%%%%%%%%%%%%%%%%%%%%%%%%%%%%%%%%%%%%

%%%%%%%%%%%%%%%%% BODY OF PAPER %%%%%%%%%%%%%%%%%%
\section{Introduction}\label{sec:intro}

The profile of emission and absorption lines in astrophysical objects encodes information about the velocity field of the gaseous medium and by extension about the physical conditions such as temperature, pressure, and depth of the gravitational potential well \citep{osterbrock89}. Emission lines of the ionized gas in the \HII\ regions in galaxies and in the narrow line regions (NLRs) around AGNs generally show single-peaked profiles due to Doppler broadening whose widths reflect the gas motion and depth of the galactic potential well \citep[e.g.,][]{whittle85,whittle85b}. Occasionally some astrophysical systems show double-peaked emission-line profiles such as those seen in the AGN NLR gas emission \citep[e.g.,][]{sargent72,heckman81,zhou04,gerke07,comerford08,xu09}. These may represent a rotating field in a disk structure similar to that seen in the protoplanetary disks surrounding Herbig Ae/Be stars \citep{acke2005}, bipolar outflows \citep[e.g., as observed in nearby Seyfert galaxies,][]{pedlar89,veilleux01,whittle04,fischer11,Wang2011a}, and/or binary orbital motion such as in double-lined spectroscopic binary stars \citep[e.g.,][]{Duquennoy1991}. 

Spatially integrated optical spectra show that 1\% of all AGNs\footnote{This fraction applies to both Type 2 \citep{Liu2010b,wang09} and Type 1 AGNs \citep{Smith2010} at least at $z\sim0.1$.} at $z\sim0.1$ exhibit double-peaked narrow emission lines with line-of-sight (LOS) velocity splitting of a few hundred km s$^{-1}$ \citep[e.g.,][]{Liu2010b,wang09,Smith2010,Ge2012}. Extensive follow-up studies based on spatially-resolved and multi-wavelength observations suggest that the velocity splitting arises from mixed origins of galactic-scale bipolar gaseous outflows, rotating galactic disks, and/or merging pairs of active supermassive black holes (SMBHs) with kpc-scale separations or ``so-called'' dual AGNs \citep[e.g.,][]{Liu2010a,Shen2011,Fu2011,Fu2011a,Fu2012,comerford11b,mcgurk11,rosario11,Peng2011,tingay11,Liu2013,Wrobel2014,Comerford2015,McGurk2015,Muller-Sanchez2015,Shangguan2016}. As such the velocity splitting in AGN narrow emission lines has been frequently adopted as a useful signpost to select candidate dual SMBHs \citep[e.g.,][]{comerford08,xu09,Liu2010b,wang09,Smith2010,popovic11,barrows12,Ge2012,Blecha2013a,Shi2014} and/or AGN-driven galactic-scale outflows \citep[e.g.,][]{greene11,Barrows2013,Zakamska2014}. 

Luminosity dependence of the frequency of AGNs with double-peaked narrow emission lines provides a useful statistical clue to the origins of the velocity splitting. If the narrow-line velocity splitting were caused by some galactic-scale extension from the radiation-pressure driven AGN outflows \citep[e.g.,][]{Arav1994,Murray1995,Proga1998,Crenshaw2003}, a positive correlation would be expected between the frequency of velocity splitting and AGN luminosity if outflows were more prevalent in more luminous AGN. Observations of the narrow UV absorbers in the broad-line region (BLR) gas suggest that the maximum BLR outflow velocity is largely set by AGN luminosity for some radiation-pressure driven outflows \citep[e.g.,][]{Perry1978,Laor2002,Vestergaard2003}. If there were some direct kinematic link between the NLR and BLR gas \citep[e.g.,][]{zamanov02}, the amplitude of velocity splitting in the NLR gas would also correlate positively with AGN luminosity \citep[e.g., see also][for the exceptionally large width and blueshifted wing observed in the \OIII\ emission in high-redshift luminous Type 1 quasars and luminous red quasars as evidence for kpc-scale ionized outflows]{Shen2015,Zakamska2015}. 

Alternatively, if the narrow-line velocity splitting arose from a rotating disk of NLR gas clouds embedded within either a stellar disk or a bulge, a weak positive correlation (or no correlation at all) of the frequency of velocity splitting with AGN luminosity would also be expected considering: i) the Tully-Fisher relation \citep{TFR} for disk galaxies or the Faber-Jackson relation \citep{Faber1976} for ellipticals; and ii) the weak correlation between AGN luminosity and galaxy mass/luminosity. The host galaxies of Type 2 AGNs are generally intermediate types with substantial stellar bulge components at least at $z\sim0.1$ \citep[e.g.,][]{kauffmann03,Greene2009,Liu2009}. Using 22,623 Type 2 AGNs with $0.02<z<0.3$ from the SDSS \citep{York2000}, \citet{kauffmann03} has shown that while more powerful AGNs are located preferentially in more massive host galaxies on average, the host stellar mass depends only very weakly on \loiii\ and that galaxies of given mass can host AGNs that span a very wide range (more than several orders of magnitude) in \loiii .

Finally, if the narrow-line velocity splitting were instead caused by the binary orbital motion of gas clouds associated with a merging pair of active SMBHs with kpc-scale separations or dual AGNs \citep[e.g., NGC 6240;][]{komossa03,max07}, some dependence on AGN luminosity would also be present. A weak positive correlation (or no correlation at all) with AGN luminosity would be expected given that galaxy mergers do enhanced AGNs as suggested by statistical studies of galaxy pairs hosting either single or double AGNs on scales smaller than 10--30 kpc \citep[e.g.,][]{ellison08,ellison11,silverman11,Liu2011a,Liu2012}. However, the effect might be too subtle to be statistically significant given that the majority of moderate luminosity AGNs is likely not associated with galaxy mergers \citep[e.g.,][]{grogin05,li06,Li2008,Liu2012}, even though merger signatures are not uncommon in AGN and in particular quasar host galaxies \citep[e.g.,][]{bahcall97,kirhakos99}. In the dual AGN scenario during the initial merging phases after the first pericenter passage, both AGN luminosity and the orbital velocities of the merging gas components would tend to increase as the merger progressed toward smaller separations, where the gravitational torque were funneling more gas to the center and more potential energy were being converted into the kinetic energy of the merging components as suggested by some simulations \citep[e.g.,][]{Blecha2013a}. On the other hand, redshift evolution might also play a role in the dual AGN scenario. For example, based on a phenomenological model, \citet{yu11} have predicted that the number of dual AGNs decreases with increasing redshift and only about 0.02\%--0.06\% of AGNs are dual AGNs with double-peaked narrow emission lines at redshifts of $z\sim$0.5--1.2.

However, most of the existing samples of AGNs with double-peaked narrow emission lines in the literature are dominated by $z\sim0.1$ galaxies and generally do not span a large enough dynamic range in AGN luminosity to study the luminosity dependence of the frequency of narrow-line velocity splitting. In this paper, we fill this gap by presenting a new sample of Type 2 AGNs with double-peaked narrow emission lines at a median redshift of $z=0.5$ with a median \OIIIb\ luminosity \loiii\ $\sim10^9L_{\odot}$ that we select from the SDSS-III/Baryon Oscillation Spectroscopic Survey \citep{Dawson2013}. Combined with the existing sample of \citet{Liu2010b} at $z\sim0.1$, the new sample has enlarged the dynamic range in \loiii\ to span almost three orders of magnitude (i.e., $10^7$--$10^{10}L_{\odot}$). With the combined sample we report the first luminosity dependence of the frequency of narrow line velocity splitting in Type II AGNs. We describe the selection and the analysis of the new sample and correction for the associated selection bias and incompleteness in \S \ref{sec:data}. In \S \ref{sec:result} we present the result on the correlation between the frequency of AGNs with double-peaked emission lines and AGN luminosity. We summarize our main findings and discuss implications of our results in \S \ref{sec:sum}.

Throughout this paper, we assume a Friedmann-Robertson-Walker cosmology with $\Omega_m = 0.3$, $\Omega_{\Lambda} = 0.7$, and $h = 0.7$. We quote velocity offsets relative to the observer, i.e., negative values mean blueshift. Following SDSS convention\footnote{https://www.sdss3.org/dr8/spectro/spectra.php}, all emission-line wavelengths are quoted in vacuum unless noted otherwise.

\begin{figure}
\centerline{
\includegraphics[width=0.23\textwidth]{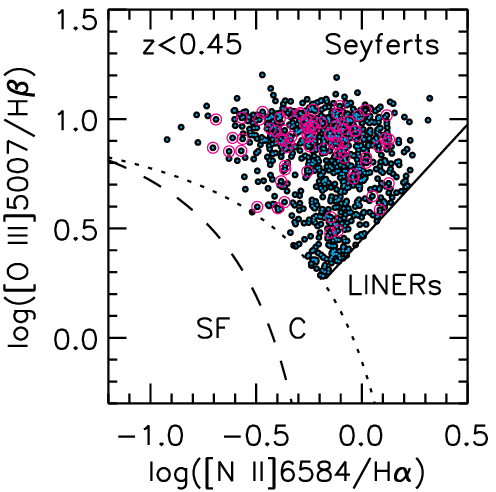} \,\,
\includegraphics[width=0.23\textwidth]{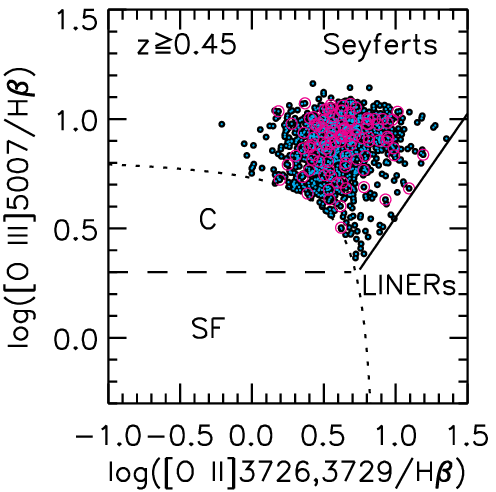}
} 
\caption{\textbf{Selection of the parent AGN sample from SDSS III/BOSS galaxies based on emission-line diagnostic diagrams.} For galaxies at $z<0.45$ (left panel), the dashed curve separates pure star-forming galaxies (``SF'') from AGN-SF composite (``C'') according to the empirical relation by \citet{kauffmann03}; the dotted curve dividing between composites and AGNs is the theoretical ``starburst limit'' from \citet{kewley01}; the solid line separating ``Seyferts'' from ``LINERs'' is from \citet{Schawinski2007}. For galaxies at $z\geq0.45$ (right panel), the dividing curves and lines are from \citet{Lamareille2010}. Our study is focused on emission-line galaxies that are classified as ``Seyferts'' for both redshift ranges. Small dots show the parent AGNs whereas large open circles in magenta indicate those selected to exhibit double-peaked narrow emission lines. See \S \ref{subsec:boss} for details.
\label{fig:bpt} }
\end{figure}

\section{Data and Method}\label{sec:data}

\subsection{The Sample from \citet{Liu2010b} at $z\sim$0.1}

To cover the relatively lower AGN luminosity regime, we adopt the sample presented by \citet{Liu2010b}. This represents a statistical sample of 167 objects with double-peaked \OIIIc\ emission lines which were selected from 14,756 type 2 AGNs with high-quality spectra drawn from the SDSS DR7 \citep{SDSSDR7}. In the general population of AGNs, emission lines such as \OIII\ are typically single-peaked and are roughly centered around or blueshifted \citep[e.g.,][]{Komossa2008d} relative to the systemic velocity of the host galaxy (as measured from stellar absorption features). In these double-peaked systems, however, one \OIII\ velocity component is redshifted and one is blueshifted from the systemic velocity of the host galaxy by a few hundred km s$^{-1}$. 

\citet{comerford09} suggested that such systems may be dual AGNs (see also \citealt{heckman81,zhou04,gerke07}), where the two \OIII\ velocity components originate from distinct NLRs around two SMBHs, co-rotating along with their own stellar bulges in a merging galaxy. Alternatively, such systems may be due to NLR kinematics such as bi-conical outflows or disk rotation \citep{duric88,axon98,crenshaw09}. Our follow up observations \citep{Liu2010a,Shen2011,Liu2013,Shangguan2016}, along with work by other groups \citep[e.g.,][]{Fu2011a,Fu2012,comerford11b,mcgurk11,Comerford2015,Shangguan2016}, have independently demonstrated that high-resolution spatial information for both \OIII\ emission and old stellar populations is key to discriminating between these alternative scenarios for individual galaxies. These individually confirmed cases of either dual AGNs or NLR kinematics also serve as a useful guide for us to address the nature for those without spatially resolved follow up observations based on their statistical properties (\S \ref{sec:sum}).

\subsection{A New Sample of Type 2 AGNs with Double-peaked Narrow Emission Lines at $z\sim0.5$}\label{subsec:boss}

To enlarge the dynamic range in AGN luminosity we supplement the \citet{Liu2010b} catalog with a new sample of 178 Type 2 AGNs with double-peaked narrow emission lines at $z\sim0.5$. Below we describe in detail the selection and properties of this new sample. 

\subsubsection{SDSS-III/BOSS Data}\label{subsubsec:boss}

We start from 933,810 galaxy spectra from the Data Release 10 \citep{SDSSDR10} of the Baryon Oscillation Spectroscopic Survey of SDSS-III \citep{Dawson2013} which was carried out using the 2.5 m telescope \citep{gunn06} at the Apache Point Observatory. We adopt stellar kinematics and emission-line flux measurements provided by the Portsmouth group \citep{Thomas2013}. This represents the largest spectroscopic sample of galaxies at a median redshift of 0.5. The galaxy sample consists of a high-redshift sample ``CMASS'' (targeting galaxies at $0.43<z<0.7$) and a low-redshift sample ``LOWZ'' (targeting galaxies at $0.15<z<0.43$). Unlike the SDSS main galaxy sample \citep{strauss02} which is largely magnitude limited, CMASS and LOWZ are both magnitude and color selected to target massive red galaxies up to redshift $z\sim0.7$ \citep{eisenstein01}. The BOSS spectrograph is 2 arcsec in fiber diameter (compared to 3 arcsec for the original SDSS spectrograph) and has an extended wavelength coverage spanning 3600 \angstrom\ to 10,000 \angstrom\ with higher throughput and a spectral resolution of $R\sim$2000. The typical signal-to-noise ratio (S/N) per \angstrom\ of a BOSS spectrum is $\sim$5 in the continua but generally much higher in the \OIII\ emission lines for the AGNs in our sample. We adopt the spectroscopic redshift determined from the BOSS pipeline and the emission-line flux measurements from \citet{Thomas2013} for our initial sample selection. As detailed below, we then use our own customized code \citep{Liu2010b} to measure the emission-line properties as presented in the analysis.

\subsubsection{AGN Selection}\label{subsubsec:agn}

\begin{figure}
\centerline{
\includegraphics[width=0.25\textwidth]{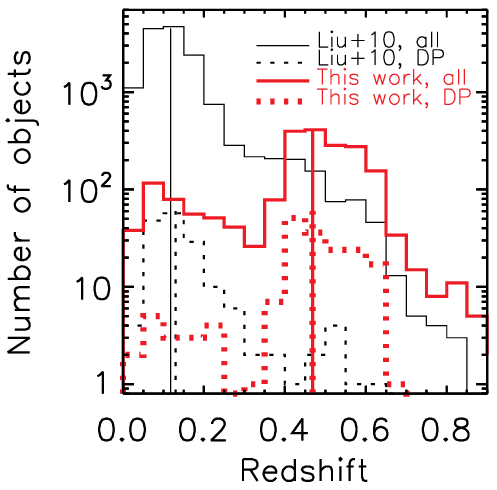}
\includegraphics[width=0.225\textwidth]{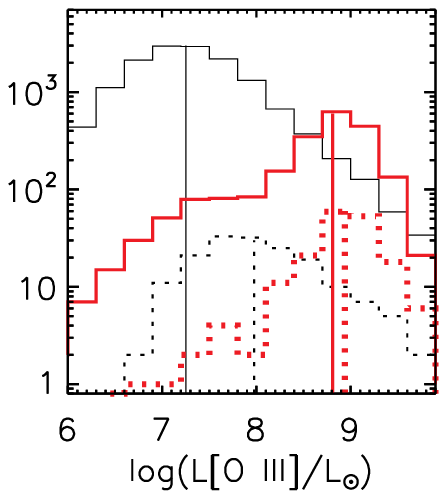}
}
\caption{\textbf{Redshift and \OIIIb\ luminosity distribution of the parent AGN sample (``all'') and those with double-peaked (``DP'') narrow emission lines.}
Vertical lines mark the median statistics of each quantity for each sample.
The new sample presented in this paper probes higher redshifts and higher \OIIIb\ luminosities on average than the \citet{Liu2010b} sample.
In particular, the new sample increases the number of Type II AGNs with double-peaked narrow emission lines (shown with dotted curves) by more than an order of magnitude at $z\sim0.5$.
} 
\label{fig:zdist}
\end{figure}

Among all the DR10 BOSS galaxies, only a small fraction has detected emission lines \citep{Thomas2013} and we focus on a high-quality subset to minimize selection biases caused by low S/N spectra. These galaxies have robust emission-line measurements with amplitude-over-noise ratio (AON) $>5$ for all the necessary emission lines relevant for AGN identification (i.e., \hbeta , \OIIIb , \halpha , and \NIIb\  for galaxies at $z<0.45$; \OIIc , \hbeta , and \OIIIb\ for galaxies at $z\geq 0.45$\footnote{We did not apply an upper-limit cut at $z=0.7$, but in practice there are only a few galaxies (50 out of 2089 or 2\%) at $0.7<z<1$ in our parent AGN sample, and no $z>0.7$ objects in our double-peaked sample (see left panel of Figure \ref{fig:zdist} and below for definition of the double-peaked sample) because the number density of BOSS galaxies drops sharply beyond that due to the flux limit of the survey \citep{White2011}. Although the ``CMASS'' sample targets galaxies at $0.43<z<0.7$ (by employing color cuts), the spectroscopically confirmed redshifts may be slightly beyond the targeted range.}). From this high-quality subset (4,835 spectra at $z<0.45$ and 2,344 spectra at $z\geq 0.45$) we select a parent sample of 2089 unique AGNs whose diagnostic emission-line ratios are classified as ``Seyfert'' with detailed selection criteria as detailed below. For AGNs with duplicate spectroscopic observations, we adopt the spectrum with the highest median S/N in our analysis.  

\begin{table*}
\begin{center}
\caption{\textbf{A new sample of 178 Type 2 Seyfert AGNs with double-peaked narrow \OIII\ emission lines at <z>$\sim$0.5 selected from SDSS-III/BOSS.}
The full table is available in the electronic version of the paper. 
The subscripts ``1'' and ``2'' stand for the blueshifted and redshifted components. 
$V_{{\rm [O\,\,III]}}$ and $V_{{\rm H\beta}}$ represent velocity offsets relative to the
systemic redshift of the host galaxy either from the BOSS pipeline (Column 5) for the 
majority of the sample or from the GANDALF-corrected redshift (Column 6) for a few objects denoted with an ``*''. 
All velocities are in units of km s$^{-1}$.
%An entry that reads zero means that the quantity is unmeasurable. 
Typical statistical errors in the FWHMs (Columns 7 \& 8) and the best-fit velocity offsets (Columns 9 \& 10) are $\sim 10$ km
s$^{-1}$ and $\sim 5$ km s$^{-1}$ for \OIII\ and are generally a few times larger for those of \hbeta .} 
\label{table:sample}
\scalebox{0.8}{
\begin{tabular}{crrrcccccccccc}
\hline
\hline
SDSS Designation & Plate & Fiber & MJD & $z_{{\rm SDSS}}$ & $z_{{\rm GANDALF}}$ & FWHM$_{{\rm [O\,\,III]},1}$ & FWHM$_{{\rm [O\,\,III]},2}$ & 
$V_{{\rm [O\,\,III]},1}$ & $V_{{\rm [O\,\,III]},2}$ &$V_{{\rm H\beta},1}$ & $V_{{\rm H\beta},2}$ & EW$_{{\rm [O\,\,III]},1}$ & EW$_{{\rm [O\,\,III]},2}$ \\
(1) & (2) & (3) & (4) & (5) & (6) & (7) & (8) & (9) & (10) & (11) & (12) & (13) & (14) \\
\hline
J000816.67+040006.6\dotfill & $4297$ & $769$ & $55806$ & $0.5740$ & $0.5737$ & $156$ & $134$ &  $-101$ &  $219$ &  $-91$ & $213$ &  $136$ & $45$  \\
J002127.73+071740.8\dotfill & $4537$ & $156$ & $55806$ & $0.2296$ & $0.2295$ & $298$ & $135$ &  $-113$ &  $264$ &  $-89$ & $234$ &  $42$ & $21$  \\
J003952.73+014829.2\dotfill & $4304$ & $302$ & $55506$ & $0.2101$ & $0.2104$ & $261$ & $169$ &  $-144$ &  $204$ &  $-159$ & $180$ &  $33$ & $14$  \\
J005021.52+073544.8\dotfill & $4543$ & $977$ & $55888$ & $0.3932$ & $0.3931$ & $178$ & $331$ & $-208$&  $123$ &  $-157$ & $117$ &  $51$ & $116$  \\
J010101.74+022640.9\dotfill & $4311$ & $595$ & $55506$ & $0.4482$ & $0.4482$ & $667$ & $118$  & $-96$  &  $166$ &  $-124$ & $188$ &  $104$ & $9$ \\
\hline
\end{tabular}}
\end{center}
\end{table*}

\begin{figure*}
  \centering
    \includegraphics[width=0.3\textwidth]{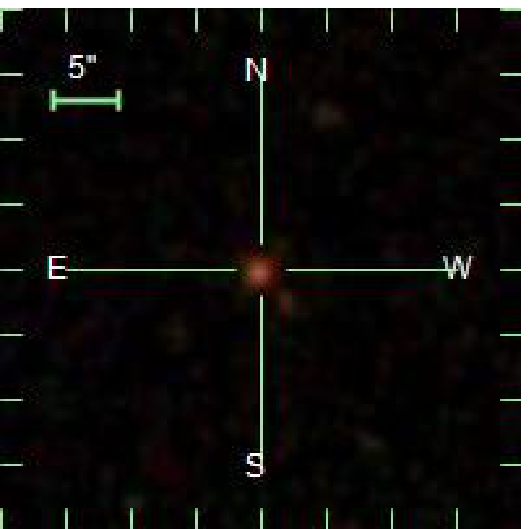}\,\,
    \includegraphics[width=0.5\textwidth]{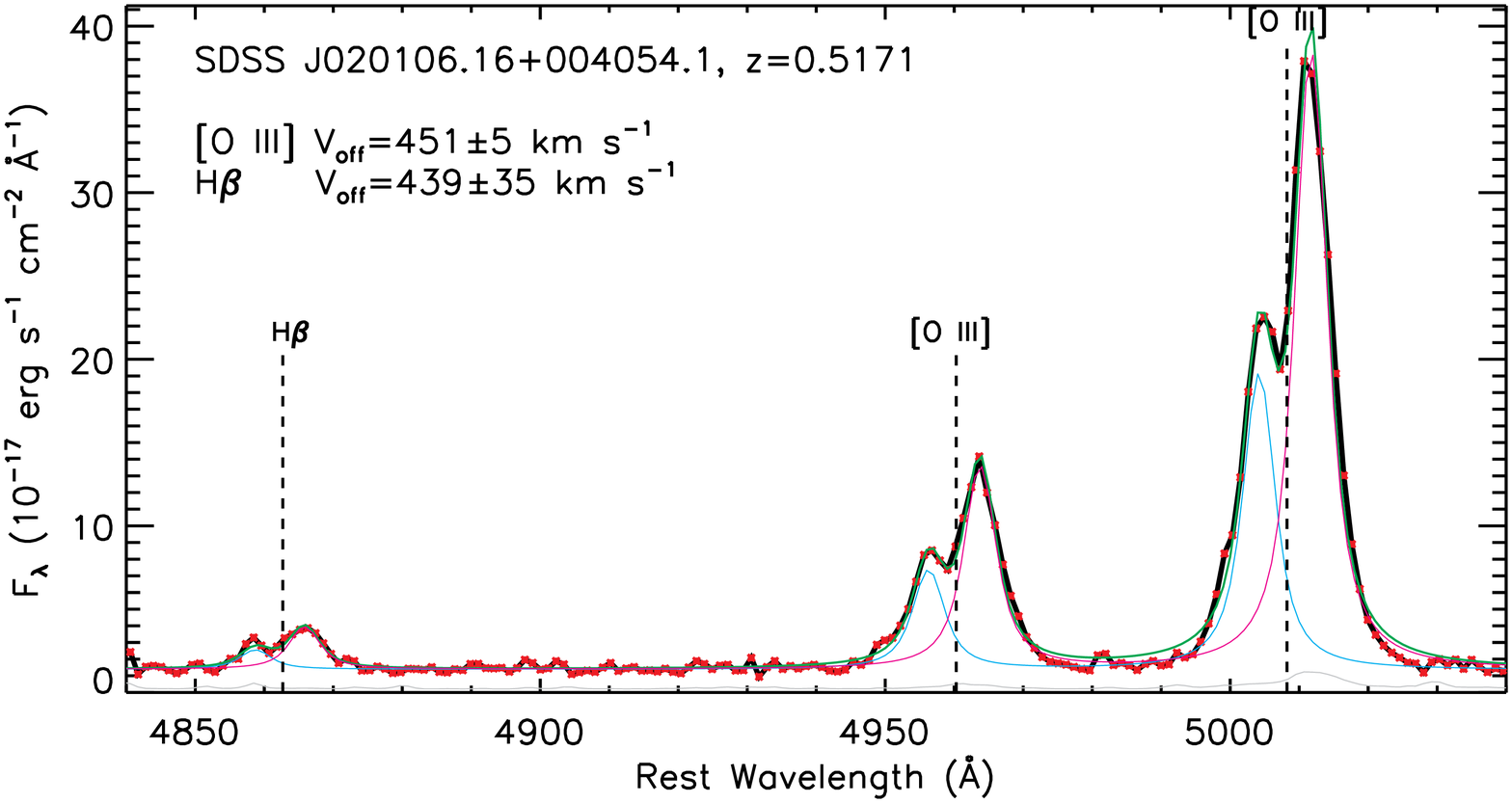}
    \includegraphics[width=0.3\textwidth]{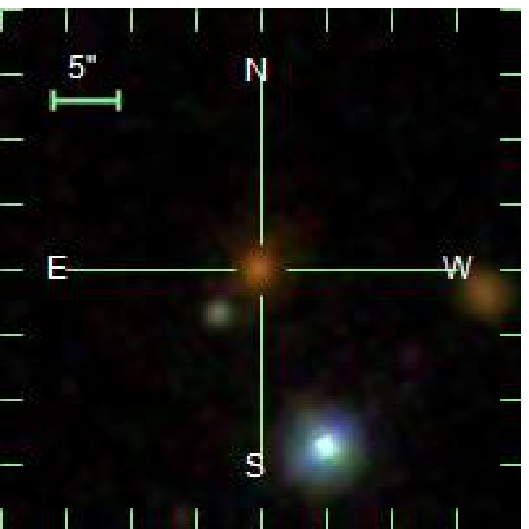}\,\,
    \includegraphics[width=0.5\textwidth]{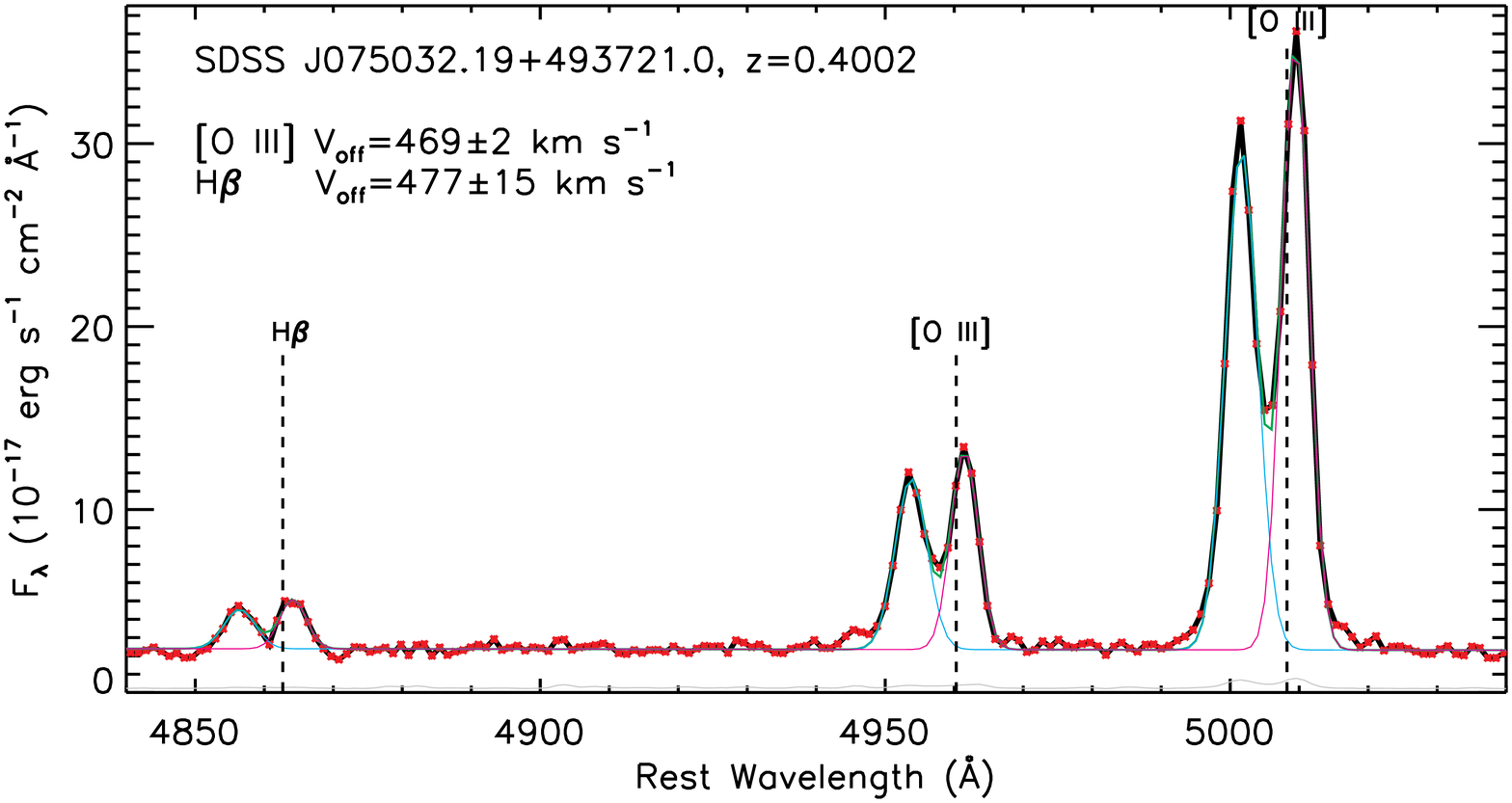}
    \includegraphics[width=0.3\textwidth]{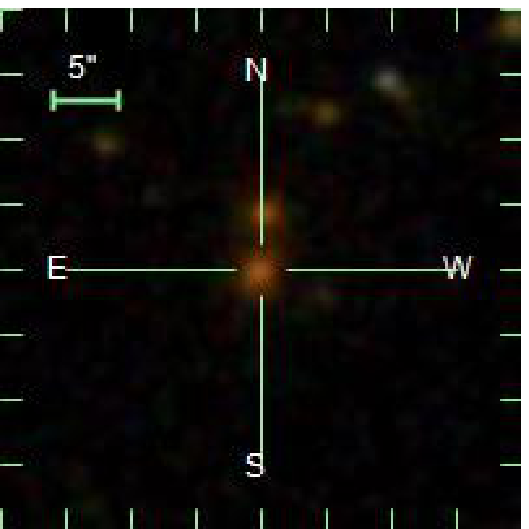}\,\,
    \includegraphics[width=0.5\textwidth]{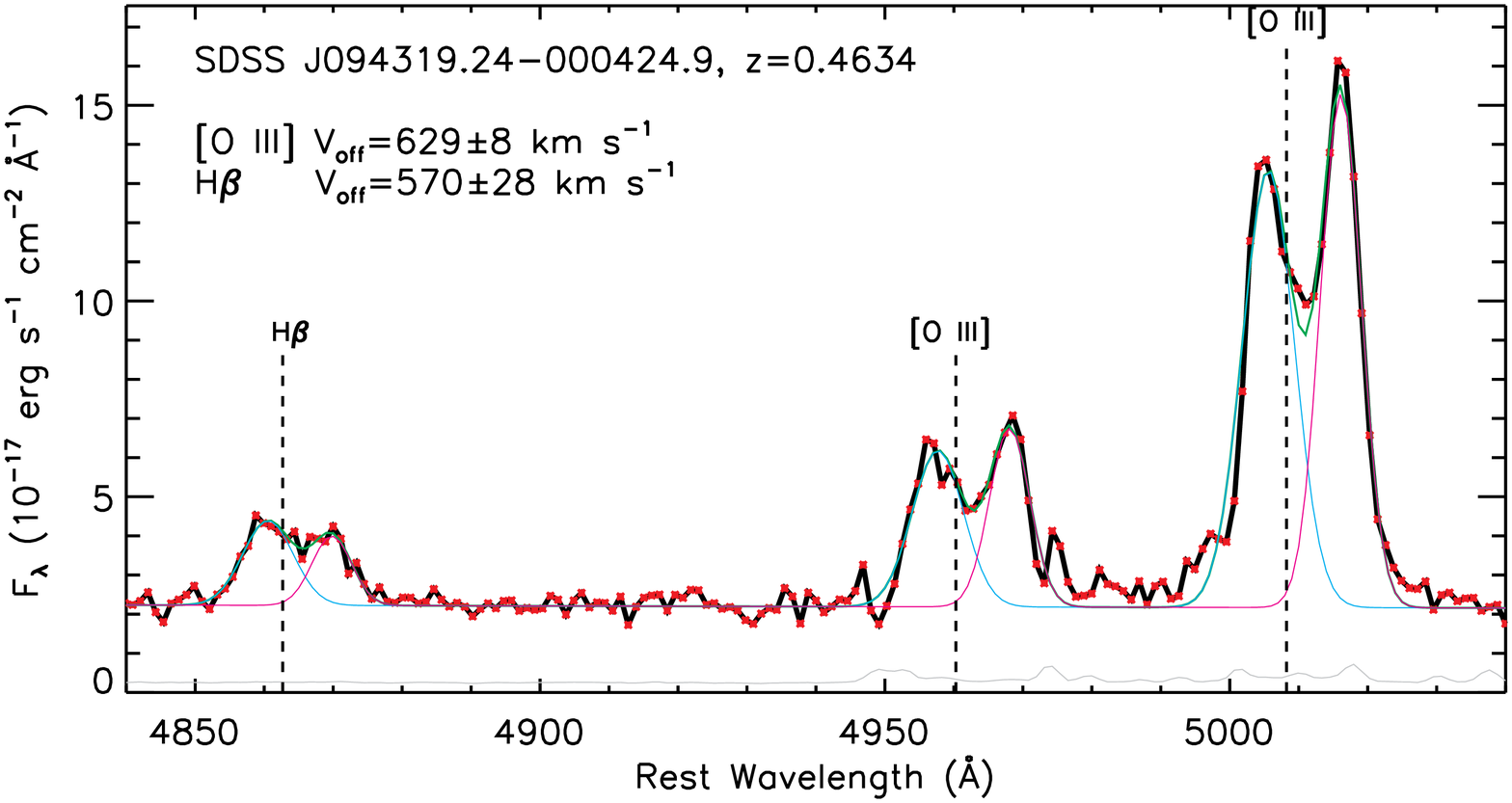}
    \caption{
   \textbf{Three examples of the sample of 178 Type 2 AGNs with double-peaked narrow \OIII\ lines newly identified from SDSS-III/BOSS.} 
    Shown here are SDSS $gri$ color-composite images (left column) and rest-frame BOSS spectra over the H$\beta$-\OIII\ region.
    (right column; data points in red and smoothed curve in black). Overplotted are our best-fit spectral models 
    (with the blueshifted and redshifted velocity components shown in cyan and in magenta, respectively, and the total in green).  
    The best-fit velocity offsets and 1-$\sigma$ statistical uncertainties between the two velocity components are shown 
    both for \OIII\ and for \hbeta\ as labeled on the plot.
    Gray curves indicate 1-$\sigma$ error spectra. The vertical lines are drawn at the
    systemic redshift of the host galaxy.}
    \label{fig:eg}
\end{figure*}

Figure \ref{fig:bpt} demonstrates the diagnostic emission-line ratios for selecting our parent AGN sample. The adopted ``Seyfert'' selection criteria are: 

\begin{enumerate}

\item For galaxies at $z<0.45$, the emission-line flux ratios \OIIIb /\hbeta\ and \NIIb /\halpha\ lie above the theoretical upper limits for star-formation excitation from \citet[][see also \citealt{Kewley2006}]{kewley01} on the BPT diagram \citep{bpt,veilleux87} and lie above the dividing line defined by \citet{Schawinski2007} to distinguish between Seyfert and low-ionization narrow emission-line regions (LINERs), the latter of which may be due to stellar/shock heating rather than AGN excitation \citep[e.g.,][]{lutz99,terashima00,ho08,Yan2012}.

\item For galaxies at $z\geq 0.45$, the emission-line flux ratios \OIIIb /\hbeta\ and \OIIc /\hbeta\ lie in the ``Seyfert'' regime (i.e., above both the star-formation and LINER regions) according to \citet{Lamareille2010}. While the major drawback of the blue emission-line diagnostic is significant mixing with star-forming objects and star-formation-AGN composites in the LINER region, it is sufficient for our purpose to select ``Seyfert'' AGNs. 

\end{enumerate}

\subsubsection{Spectral Modeling and Identification of Double Peaks}\label{subsubsec:dp}

We model the continuum and emission lines adopting the same procedure of \citet{Liu2010b} as we summarize below. 
First, we fit the galaxy continuum over emission-line-free regions using a best-fit template that we construct from a linear combination of instantaneous starburst models of \citet{bc03} following the method described in \citet{Liu2009}. For the majority of the sample AGNs whose host galaxy stellar continua are too weak (median continuum S/N $<$ 10 pixel$^{-1}$) for accurate stellar template fitting, we adopt a simple power-law model for the continuum instead. 

Second, we fit the continuum-subtracted \OIII\ region over a rest-frame wavelength range of $\lambda\lambda$4930--5040\angstrom\ using a double-Lorentzian model constructed by a pair of Lorentzian functions with different velocities convolved with the measured instrumental resolution of the BOSS spectra ($\sigma_{{\rm ins}}\sim$ 65 km s$^{-1}$) for each of the \OIIIa\ and \OIIIb\ lines. For each velocity component of \OIIIa\ and \OIIIb , both the redshift and line width are constrained to be the same. We allow the flux ratio to vary between \OIIIa\ and \OIIIb , although it is always close to 3. 

Third, we also model the \hbeta\ region ($\lambda\lambda$4850--4880\angstrom ) with a double-Lorentzian model of which the line widths of both components are fixed to be the same as those of \OIII\ whereas the velocities are allowed to be different from those of \OIII . 
Fourth, we re-do the fitting with a double-Gaussian model instead, and take the one with a smaller reduced $\chi^2$ as the best-fit model under the two-velocity-component assumption. We also re-do the fitting assuming that the emission lines are all single peaks by using either a single-Lorentzian or single-Gaussian model to compare with the two-velocity-component models. 

Finally, we identify 178 systems with double-peaked narrow emission lines for which both \OIIIa\ and \OIIIb\ are better modeled with two velocity components rather than the one-component models. In all the cases where \hbeta\ is measurable, it also exhibits a double-peaked profile whose velocity offset is similar to that of \OIII . We have verified the selection based on model fitting by visual inspection of all the spectra. Same as in \citet{Liu2010b}, we focus on AGNs with well-detected double peaks in both \OIIIa\ and \OIIIb\ with consistent profiles; our sample generally does not include more complex profiles such as lumpy, winged, or multi-component velocity signatures. As shown in Figure \ref{fig:bpt}, the double-peaked sample spans the full parameter space of the ``Seyfert'' population occupied by the parent AGN sample on the emission-line diagnostic diagrams.

Figure \ref{fig:zdist} shows the systemic redshift and \OIIIb\ luminosity distribution of the new sample of 178 AGNs that we select with double-peaked \OIII\ lines as well as that of the parent AGN sample. Also shown for comparison is the $z\sim0.1$ sample of \citet{Liu2010b}. The new sample probes higher redshifts (median $z\sim0.5$ compared to median $z\sim0.1$) and by extension higher luminosities (median \loiii\ $\sim10^9L_{\odot}$ compared to median \loiii\ $\sim10^8L_{\odot}$) on average than the \citet{Liu2010b} sample.  In particular, the new sample increases the known number of Type 2 AGNs with double-peaked narrow emission lines by more than an order of magnitude at $z\sim0.5$.

\begin{figure}
\centerline{
\includegraphics[width=0.5\textwidth]{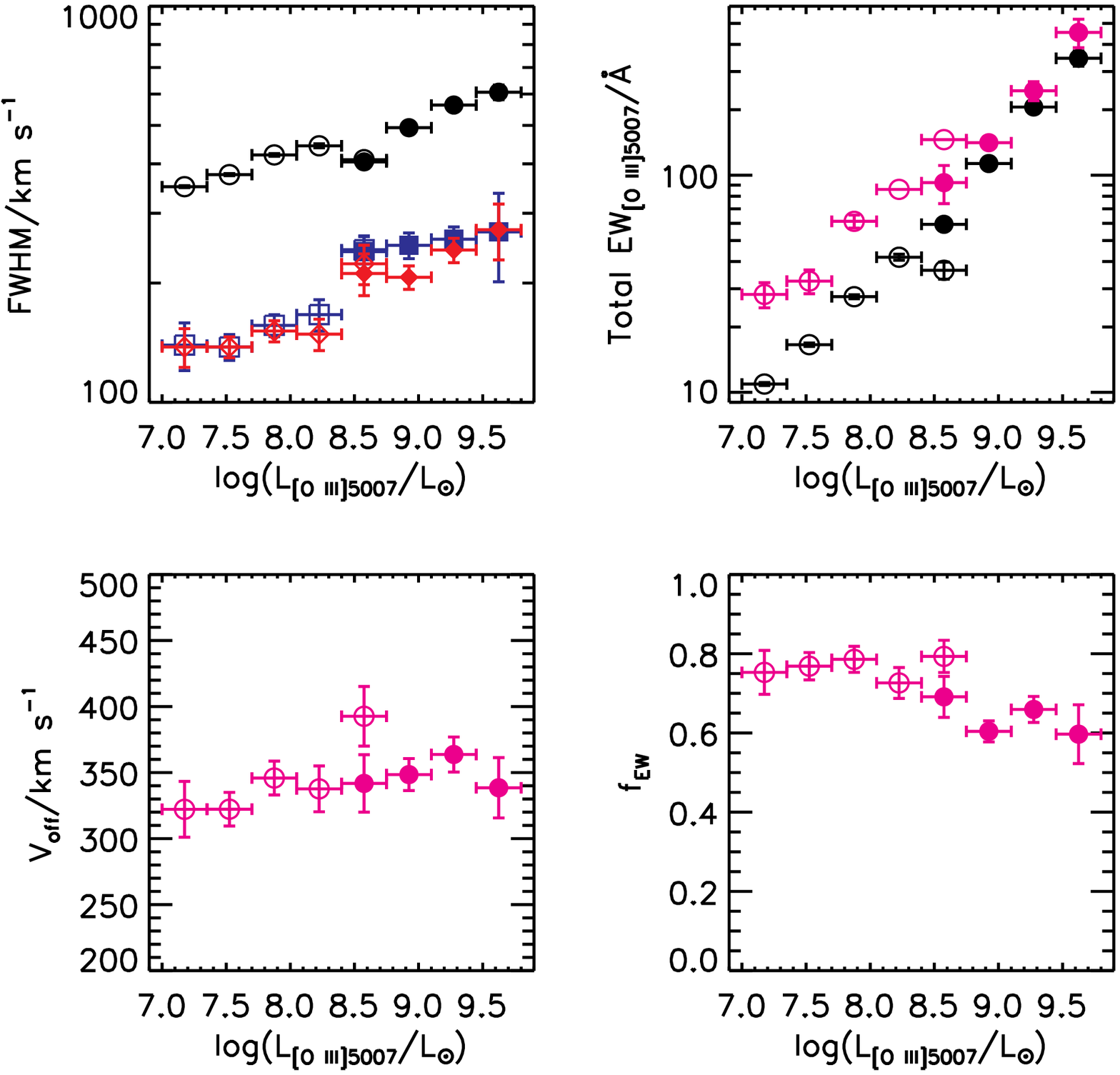}
} \caption{\textbf{Correlations between the total \OIII\ luminosity \loiii\ and FWHM, total \OIII\ EW, velocity offset, and the flux ratio between the two velocity components.}
The points shown represent median values of the parameters for all the objects at each given bin of \OIII\ luminosity. Error bar denotes uncertainty in the median value, which was estimated using the standard deviation divided by the square root of the number of objects in any given luminosity bin.
Filled symbols represent measurements of the new sample whereas open symbols denote those based on the \citet{Liu2010b} sample. 
Color symbols denote measurements of the double-peaked objects (with red for redshifted velocity component and blue for blueshifted component; magenta for measurements based on both velocity components) whereas black symbols are those from the parent AGN sample. 
The correlations between \loiii\ and these various emission-line properties (in particular with the FWHM and total EW) introduce luminosity-dependent selection biases 
that have been corrected for (\S \ref{subsubsec:bias}) to study the intrinsic luminosity dependence on the fraction of double-peaked emission lines.}
\label{fig:bias} 
\end{figure}

Figure \ref{fig:eg} shows the SDSS images \citep{lupton04} and BOSS spectra for three individual examples of AGNs in the new sample of 178 Type 2 AGNs with double-peaked \OIII\ emission lines. Table \ref{table:sample} lists their SDSS names (with RA and Dec given in the J2000 coordinates), systemic redshift of the host galaxy from the SDSS SpecBS pipeline, GANDALF-corrected redshift, SDSS spectroscopic plate number, fiber ID, MJD of the observation date, as well as their emission-line kinematic properties from our spectral analysis including the LOS velocity offsets of the double-peaked components in either \OIII\ or \hbeta\ and the rest-frame EW of the individual \OIII\ peak. The GANDALF-corrected redshift in the Portsmouth catalog \citep{Thomas2013} was based on the publicly available code GANDALF \citep{Sarzi2006} which simultaneously fits stellar population and Gaussian emission line models to the galaxy spectrum. We use the difference between the redshift from the SDSS pipeline and the GANDALF-corrected redshift as a measure of the systematic uncertainty in the systemic redshift of the host galaxy. We have adopted the BOSS pipeline redshift as our baseline values for the majority of the sample, although for 8 out of the 178 objects (denoted with an ``*'' in Table \ref{table:sample}) we have adopted the GANDALF-corrected redshift instead as our baseline value which better fits the observed spectrum\footnote{The exact systemic redshift adopted does not affect our main result on the relative LOS velocity offset between the blueshifted and redshifted velocity components in the emission lines. We have also verified that adopting the SDSS pipeline redshift also for the 8 objects would not change the results significantly either for the analysis involving velocity offsets relative to the host-galaxy systemic redshift (see discussion in \S \ref{sec:sum}), although adopting the GANDALF-corrected redshift for all the 178 objects in the double-peaked sample, on the other hand, would introduce significant more scatter than using the SDSS pipeline redshift for the results involving velocity offsets relative to the host-galaxy systemic redshift}. Hereafter we adopt velocity offsets measured based on \OIII\ in our analysis which is more robust than that based on \hbeta . 

\subsubsection{Correcting for Selection Bias and Incompleteness}\label{subsubsec:bias}

Using Monte Carlo simulations, \citet{Liu2010b} has demonstrated that the relative selection completeness of double-peaked narrow line objects is a strong\footnote{The relative completeness has little dependence on the median S/N of the spectrum, because the selection is mainly affected by emission lines and not by continuum.} function of emission-line width (as quantified by FWHM), line strength (characterized by \OIII\ EW), velocity offset $v_{{\rm off}}$, and the flux ratio between the two velocity components $f_{{\rm EW}}$. AGNs with smaller \OIII\ FWHM, larger \OIII\ EW, larger $v_{{\rm off}}$, and larger $f_{{\rm EW}}$ are more likely to be selected to have double-peaked \OIII\ emission lines. Any correlation between AGN luminosity and these parameters would therefore introduce luminosity-dependent selection bias and incompleteness which we must correct for in order to address the intrinsic luminosity dependence of the double-peaked narrow line fraction.

Figure \ref{fig:bias} demonstrates the relations between \OIII\ luminosity and FWHM, \OIII\ EW (for both the double-peaked and the parent samples), $v_{{\rm off}}$, and $f_{{\rm EW}}$ for our new BOSS double-peaked sample as compared to the \citet{Liu2010b} sample. The points shown represent median values of the parameters for all the objects at each given bin of \OIII\ luminosity. Error bar denotes uncertainty in the median value, which was estimated using the standard deviation divided by the square root of the number of objects in any given luminosity bin. The correlation is the strongest for \OIII\ EW and weaker for FWHM, and there is little correlation with $v_{{\rm off}}$ and $f_{{\rm EW}}$. Based on these observed relations and the relative selection completeness determined from our Monte Carlo simulations \citep{Liu2010b}, we correct for selection bias and incompleteness by estimating the relative correction factors which are then divided out to infer the {\it intrinsic} luminosity dependence for the fraction of double-peak narrow emission lines\footnote{More specifically, our Monte Carlo simulations provide a relative selection completeness factor as a function of each of the four parameters \citep[i.e., FWHM, \OIII\ EW, $v_{{\rm off}}$, and $f_{{\rm EW}}$;][]{Liu2010b}. Each of the four parameters in turn is a function of \OIII\ luminosity (Figure \ref{fig:bias}). For at any given \OIII\ luminosity bin, we first infer the median value of the parameter from the correlations as shown in Figure \ref{fig:bias}, and then estimate the relative selection completeness factor for that particular parameter as determined by simulations. Finally we multiply the relative selection completeness factors for all the four parameters together as the total completeness factor at a given \OIII\ luminosity bin.} (\S \ref{sec:result}). We caution, however, that the {\it absolute} completeness depends on the actual underlying distributions of all the relevant emission-line properties. Nevertheless, the completeness estimates are useful to correct for {\it relative} selection bias to properly address luminosity dependence.

\section{Result: Luminosity Dependence of AGN Narrow-line Velocity Splitting}\label{sec:result}

\begin{figure}
\centerline{
\includegraphics[width=0.45\textwidth]{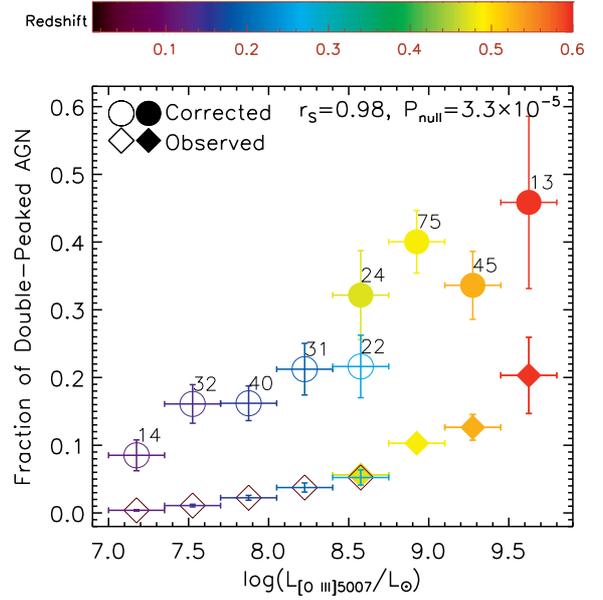}
} 
\caption{\textbf{Luminosity dependence of the relative fraction of AGNs with double-peaked narrow emission lines before (diamonds) and after correction (circles) for selection bias and incompleteness (\S \ref{subsubsec:bias}).} We have estimated the relative correction factors based on: a) the observed relations between \OIII\ luminosity and the relevant emission line properties (Figure \ref{fig:bias}); and b) the relative selection completeness as determined from Monte Carlo simulations \citep{Liu2010b}. The corrected fraction is only appropriate in a {\it relative} sense in that we do not know the true underlying distribution of the properties of double-peaked narrow emission lines. Labeled on the plot are the Spearman correlation coefficient $\rho$ and the null probability for the corrected fraction. 
Open circles are from \citet{Liu2010b} whereas filled circles are from the new sample presented in this paper. 
The new sample on average probes systematically higher \OIII\ luminosity, enabling a larger enough dynamic range to study the luminosity dependence. 
Color encodes the median redshift of the parent galaxies within each luminosity bin. 
No significant redshift evolution ($<2\sigma$) is detected in our sample when controlled in AGN luminosity.
We label the numbers of the double-peaked objects contained in each bin. 
Only bins with more than 10 double-peaked objects are shown. 
Error bars denote 1 $\sigma$ Poisson errors. 
See \S \ref{subsec:frac} for more details.
\label{fig:deplo3} }
\end{figure}

\subsection{The Fraction of Type 2 AGNs with \OIII\ Double Peaks}\label{subsec:frac}

Figure \ref{fig:deplo3} shows the dependence of the fraction of double-peaked narrow emission lines among all the parent sample of Type 2 AGNs classified as Seyferts (\S \ref{subsec:boss}) on AGN luminosity as indicated by \OIII\ luminosity\footnote{We do not use Eddington ratio because our targets are Type 2 AGNs without direct estimates on the black hole masses \citep[e.g.,][]{Shen2013} and that indirect black-hole mass estimators such as the host-galaxy stellar velocity dispersion, $\sigma_{*}$ \citep[e.g.,][]{tremaine02}, is only available for a small fraction of our double-peak sample (see discussion below in \S \ref{sec:sum}), even though that both black hole mass and Eddington ratio are known to be better indicators for the underlying physical processes in AGNs \citep[e.g.,][]{laor00,Boroson2002,Shen2014}.} \loiii . The apparent double-peaked narrow line fraction (shown with diamonds) increases from $1\pm0.1$\% (1 $\sigma$ Poisson errors) at \loiii\ $=10^7L_{\odot}$ to $10\pm2$\% at \loiii\ $=10^{9.0}L_{\odot}$ and $20\pm6$\% at \loiii\ $\gtrsim10^{9.5}L_{\odot}$. The overall double-peaked fraction observed for our new BOSS sample is $\sim9\pm1$\% (or 178 out of 2089 AGNs), which is almost an order of magnitude larger than that seen in the $z\sim0.1$ samples taken at face value (e.g., $\sim1.1\pm0.1$\% or 167 out of 14756 AGNs, \citealt{Liu2010b}).

Figure \ref{fig:deplo3} also shows the intrinsic {\it relative} \loiii\ dependence (shown with circles) of the double-peaked narrow-line fraction after correcting for the selection bias and incompleteness due to the luminosity dependence of the various relevant emission-line properties (Figure \ref{fig:bias}). The correction accounts for the fact that the identification of narrow-line double peaks is more incomplete for AGNs with larger \OIII\ FWHM, smaller \OIII\ EW (and therefore low S/N), smaller velocity offset, and smaller line flux ratio. The corrected intrinsic luminosity dependence of the double-peaked fraction becomes weaker but is still statistically significant (Spearman's rank correlation coefficient $\rho=0.98$ with a null probability of $P_{{\rm null}}=3.3\times10^{-5}$, or effectively a $\sim4.2\sigma$ detection for a positive correlation). The corrected relative fraction increases from $8\pm2$\% at \loiii\ $=10^7L_{\odot}$ to $40\pm5$\% at \loiii\ $=10^{9.0}L_{\odot}$ and $46\pm13$\% at \loiii\ $\gtrsim10^{9.5}L_{\odot}$.

No significant redshift evolution ($<2\sigma$) is detected in our sample when controlled in AGN luminosity, although the dynamic range in redshift is relatively small and our sample is highly incomplete for faint AGNs at high redshift due to the flux limit of the BOSS survey.

\subsection{Maximum Splitting Velocity}\label{subsec:vmax}

\begin{figure}
\centerline{
\includegraphics[width=0.45\textwidth]{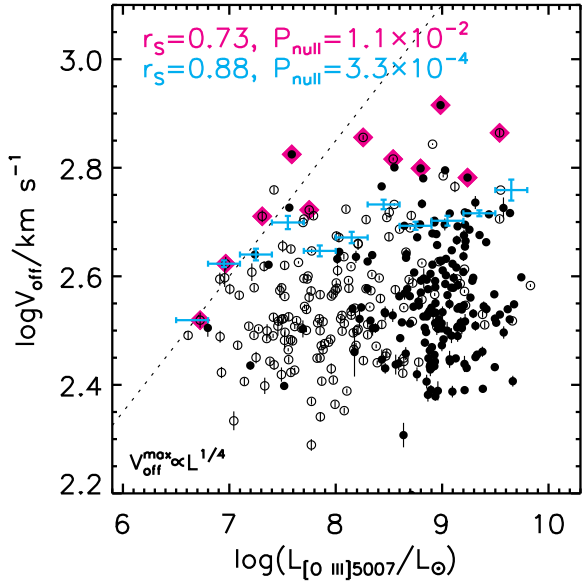}
} 
\caption{\textbf{Relation between \OIII\ luminosity and the LOS splitting velocity $V_{{\rm off}}$ between the blueshifted and redshifted \OIII\ components.}
We quantify the maximum LOS splitting velocity, $V^{{\rm max}}_{{\rm off}}$, using either the largest splitting velocity for all AGNs in a given luminosity bin (shown as diamonds in magenta), or the medium value of all the objects in the top 20 percentile in splitting velocity in a given luminosity bin (shown in cyan with error bars representing 1 $\sigma$ Poisson errors).
Indicated on the plot are Spearman correlation coefficient $\rho$ and the null probability for each version of the estimated $V^{{\rm max}}_{{\rm off}}$. 
Small open circles are from \citet{Liu2010b} whereas filled ones are from the new sample. 
Error bars on $V_{{\rm off}}$ denote 1 $\sigma$ statistical errors most of which are smaller than the symbol size. 
The dotted line marks the theoretical curve as expected for some radiation-pressure driven outflows 
with $V^{{\rm max}}_{{\rm off}}\propto L^{1/4}$ normalized at the lowest luminosity bin.
See \S \ref{subsec:vmax} for more details.
}
\label{fig:vmax} 
\end{figure}

Figure \ref{fig:vmax} shows the LOS splitting velocity $V_{{\rm off}}$ between the blueshifted and redshifted \OIII\ components as a function of \OIII\ luminosity \loiii . To quantify the maximum LOS splitting velocity $V^{{\rm max}}_{{\rm off}}$ at given \loiii , we have examined two surrogates: i) the largest $V_{{\rm off}}$ for all objects in a given luminosity bin, and ii) the medium $V_{{\rm off}}$ of objects in the top 20 percentile of $V_{{\rm off}}$ in each luminosity bin. The former provides a more accurate representation of $V^{{\rm max}}_{{\rm off}}$ whereas the latter is less sensitive to outliers due to small number statistics. In the latter case $V^{{\rm max}}_{{\rm off}}$ is correlated with \loiii\ (with Spearman's rank correlation coefficient $\rho=0.88$ and $P_{{\rm null}}=3.3\times10^{-4}$ or $\sim3.5\sigma$) whereas there is no significant correlation (Spearman's rank correlation coefficient $\rho=0.73$ and $P_{{\rm null}}=1.1\times10^{-2}$ or $<2.5\sigma$) between $V^{{\rm max}}_{{\rm off}}$ and \loiii\ in the former case which is more subject to outliers due to small number statistics.

\begin{figure*}
\centerline{
\includegraphics[width=0.99\textwidth]{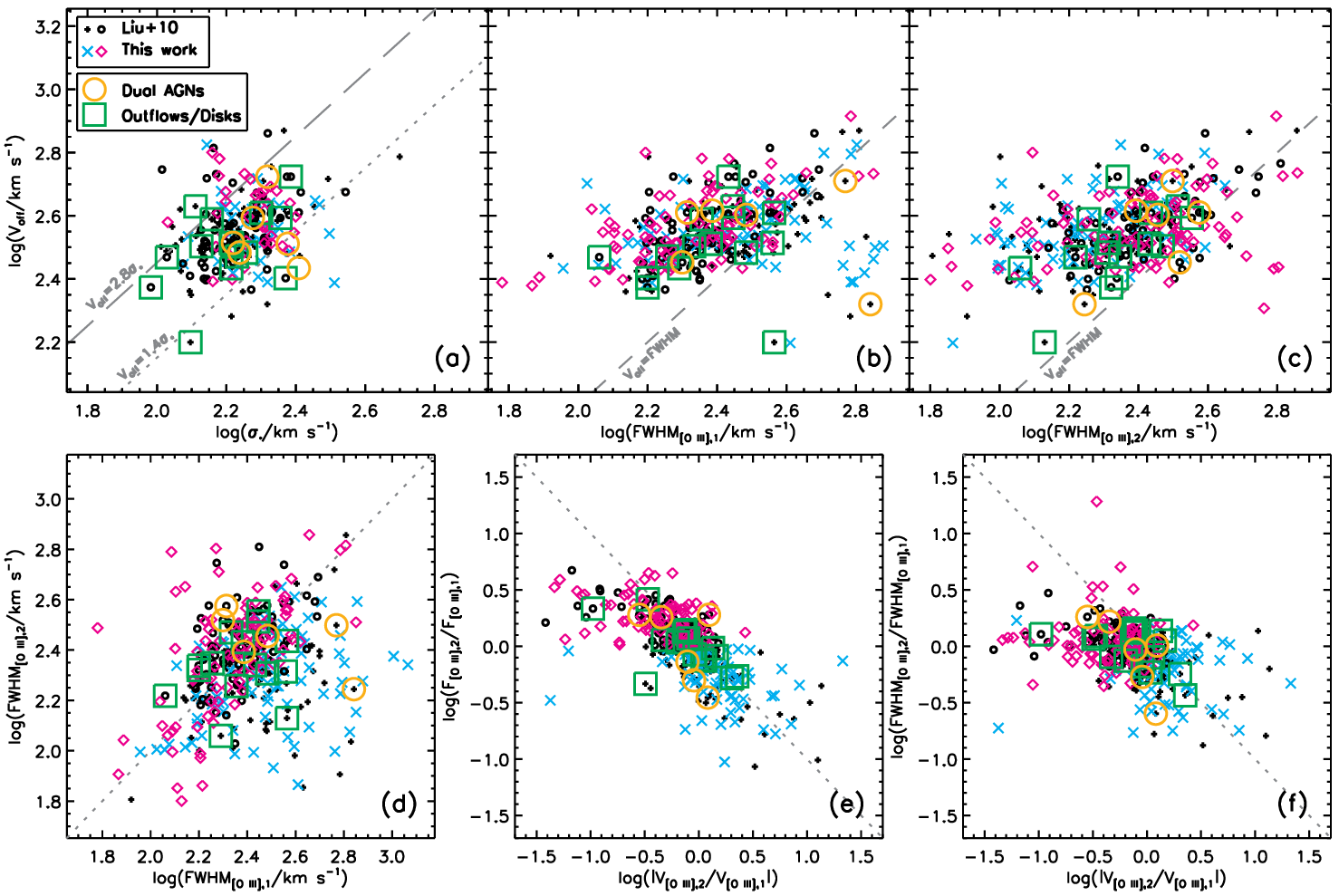}
} 
\caption{\textbf{Dynamical properties of the double-peaked \OIIIb\ components.} 
We compare our new sample from SDSS III/BOSS at $z\sim0.5$ (``This work'') with the \citet{Liu2010b} sample (``Liu+10'') at $z\sim0.1$. 
Objects with more luminous blueshifted velocity components are plotted as crosses in cyan for the new sample and plus signs in black for the \citet{Liu2010b} sample.
Objects with more luminous redshifted velocity components are plotted as open diamonds in magenta for the new sample and small open circles in black for the \citet{Liu2010b} sample. Also indicated with large symbols are the few known cases of dual AGNs and NLR outflows and/or rotating disks from the literature \citep{Liu2010a,Shen2011,Fu2012} which were identified from the \citet{Liu2010b} sample using spatially resolved followup observations.
Refer to Table \ref{tab:stat} for results on the correlation test of each relation for each sample.
(a). stellar velocity dispersion $\sigma_{*}$ vs. velocity offset between the double-peaked \OIII\ components.
Only objects with robust $\sigma_{*}$ measurements (S/N$>5$) are shown.
(b). FWHM of the blueshifted \OIII\ component vs. velocity offset between the double-peaked \OIII\ components.
(c). FWHM of the redshifted \OIII\ component vs. velocity offset between the double-peaked \OIII\ components.
(d). FWHM of the blueshifted \OIII\ component vs. that of the redshifted \OIII\ component.
(e). Velocity-offset ratio of the double-peaked \OIII\ components vs. the \OIII\ line flux ratio.
(f). Velocity-offset ratio of the double-peaked \OIII\ components vs. the \OIII\ line FWHM ratio.
See \ref{sec:sum} for more discussion.
}
\label{fig:dynamics} 
\end{figure*}

\section{Summary and Discussion}\label{sec:sum}

We have studied the luminosity dependence of the fraction of double-peaked narrow emission lines in optically selected Type 2 (obscured) AGNs using \OIIIb\ luminosity as a surrogate for AGN luminosity.  We have combined the sample of \citep{Liu2010b} at $z\sim0.1$ with a new sample of 178 AGNs at $z\sim0.5$ with double-peaked narrow lines that we have identified from the SDSS-III/BOSS survey. The new sample was selected from a parent sample of 2089 Type 2 AGNs with high quality BOSS spectra (amplitude-over-noise ratios $>5$ for all the relevant emission lines). It includes more galaxies at higher redshifts with larger intrinsic AGN luminosity which allows us to probe a large enough dynamic range to address the luminosity dependence. We summarize our main conclusions as the following. 

\begin{enumerate}

\item We have found a statistically significant positive correlation ($P_{{\rm null}}=3.3\times10^{-5}$ or $\sim4.3\sigma$) between AGN \OIIIb\ luminosity \loiii\ and the fraction of double-peaked narrow lines after correcting for selection bias due to \OIII\ equivalent width, line width, velocity splitting, and/or flux ratio. The apparent fraction increases from $1\pm0.1$\% (1 $\sigma$ Poisson errors) at \loiii\ $=10^7L_{\odot}$ to $10\pm2$\% at \loiii\ $=10^{9.0}L_{\odot}$ and $20\pm6$\% at \loiii\ $\gtrsim10^{9.5}L_{\odot}$, whereas the corrected relative fraction increases from $8\pm2$\% at \loiii\ $=10^7L_{\odot}$ to $40\pm5$\% at \loiii\ $=10^{9.0}L_{\odot}$ and $46\pm13$\% at \loiii\ $\gtrsim10^{9.5}L_{\odot}$. 

\item We have also found tentative evidence for a positive correlation between \loiii\ and the maximum LOS splitting velocity $V^{{\rm max}}_{{\rm off}}$ between the blueshifted and redshifted components of \OIII . The correlation is statistically significant ($\sim3.5\sigma$) if we quantify $V^{{\rm max}}_{{\rm off}}$ approximately using the median value of the offset velocity of the top 20 percentile of all AGNs in a given luminosity bin.  No significant correlation ($<2.5\sigma$) is found if we directly use the largest velocity offset for all AGNs in a given luminosity bin which, however, is more sensitive to outliers due to small number statistics. 

\end{enumerate}

The apparent rapid increase of the fraction of AGNs with double-peaked narrow emission lines with increasing luminosity at \loiii\ $>10^{8.5}L_{\odot}$ is not surprising considering the luminosity-dependent selection bias (Figure \ref{fig:bias}). Previously \citet{Liu2010b} also reported evidence for a factor of 2 increase from \loiii\ $=10^{7}L_{\odot}$ to \loiii\ $=10^{8.5}L_{\odot}$ after correcting for selection bias and incompleteness. The large fraction of double peaks at the highest luminosities may indicate that the majority of the most luminous AGNs trigger outflows and/or host mergers of SMBHs, after accounting for the fact a substantial fraction would be missed as single peaks due to projection effects. As we discuss below, our results suggest that more luminous AGNs are more likely to drive galactic-scale outflows and/or host merging massive black holes. 

% ---------------------------------------------
\begin{table*}
\begin{center}
\caption{\textbf{Results from Spearman correlation tests on the various dynamical properties of the double-peaked components as shown in Figure~\ref{fig:dynamics} for AGNs with double-peaked \OIII\ emission lines from the new sample presented in this work (Columns 4--6) as compared with the \citet{Liu2010b} sample (``Liu+10''; Columns 7--9).}
Subset A stands for objects with a more luminous blueshifted velocity component than the redshifted component (as would be expected for an intrinsically symmetric outflow after accounting for the effect of dust extinction) whereas Subset B represents those with a more luminous redshifted component than the blueshifted one.
``All DP'' denotes the whole sample of double-peaked objects, i.e., the combination of Subset A and Subset B for each case. 
Numbers shown in brackets are 1-$\sigma$ errors estimated from Monte Carlo simulations using Bootstrap resampling.
}
\label{tab:stat}
\scalebox{0.78}{
\begin{tabular}{ccccccccc}
\toprule
\toprule
   & & & & This work & & & Liu+10 &  \\
    \cmidrule(lr){4-6} \cmidrule(lr){7-9}
x & y & Statistics\dotfill & Subset A & Subset B & All DP & Subset A & Subset B & All DP \\
(1) & (2) & (3) & (4) & (5) & (6) & (7) & (8) & (9) \\
\hline
%%%%%%%%%%%%%%%%%%%%%%%%%%%%%%%%%%%%%%%%%%%%%%%%%% 
$\sigma_{\ast}$ & $V_{{\rm off}}$ & $\rho$\dotfill & 0.0 & $-0.1$ & 0.0 & 0.4 & 0.4 & 0.4 \\
 & & & [$-0.2$, $0.2$] & [$-0.2$, $0.1$]  & [$-0.2$, $0.1$] & [0.3, 0.5] & [0.3, 0.5] & [0.4, 0.5] \\
 & & $P_{{\rm null}}$\dotfill & 1.0 & 0.8 & 0.9 & $3\times10^{-4}$ & $9\times10^{-5}$& $3\times10^{-8}$ \\
 & & & [0.1, 1.0] & [0.1, 0.8] & [0.1, 0.9] & [$3\times10^{-6}$, $1\times10^{-2}$] &[$6\times10^{-7}$, $7\times10^{-3}$]  & [$5\times10^{-11}$, $9\times10^{-6}$] \\
\cmidrule(lr){4-6} \cmidrule(lr){7-9}
%%%%%%%%%%%%%%%%%%%%%%%%%%%%%%%%%%%%%%%%%%%%%%%%%% 
FWHM$_{\rm [O\,\,III],1}$ & $V_{{\rm off}}$ & $\rho$ \dotfill & 0.2 & 0.5 & 0.4 & 0.3 & 0.6 & 0.4 \\
& & & [$0.0$, $0.3$] & [$0.5$, $0.6$] & [$0.3$, $0.4$] &[$0.2$, $0.4$]  & [$0.5$, $0.7$] & [$0.3$, $0.5$] \\
         &  & $P_{{\rm null}}$  \dotfill & 0.1 & $1\times10^{-8}$ & $4\times10^{-7}$ & $7\times10^{-3}$ & $4\times10^{-9}$ & $3\times10^{-8}$ \\
 & & & [$8\times10^{-3}$, 0.6] & [$1\times10^{-11}$, $2\times10^{-6}$] & [$3\times10^{-10}$, $7\times10^{-5}$] & [$4\times10^{-5}$, $8\times10^{-2}$] & [$8\times10^{-12}$, $1\times10^{-6}$] & [$8\times10^{-12}$, $4\times10^{-6}$] \\
\cmidrule(lr){4-6} \cmidrule(lr){7-9}
%%%%%%%%%%%%%%%%%%%%%%%%%%%%%%%%%%%%%%%%%%%%%%%%%% 
FWHM$_{\rm [O\,\,III],2}$ & $V_{{\rm off}}$ & $\rho$ \dotfill & 0.4 & 0.3 & 0.3 & 0.4 & 0.5 & 0.5 \\
 & & & [0.3, 0.5] & [0.2, 0.4] & [0.3, 0.4] & [0.3, 0.5] & [0.4, 0.6] & [0.4, 0.5] \\
 & & $P_{{\rm null}}$\dotfill & $1\times10^{-4}$ & $4\times10^{-3}$ & $5\times10^{-6}$ & $2\times10^{-4}$ & $2\times10^{-7}$ & $7\times10^{-10}$ \\
 & & & [$1\times10^{-6}$, $5\times10^{-3}$] & [$6\times10^{-5}$, $6\times10^{-2}$] & [$2\times10^{-8}$, $4\times10^{-4}$] & [$3\times10^{-6}$, $1\times10^{-2}$] & [$6\times10^{-10}$, $3\times10^{-5}$] & [$1\times10^{-12}$, $6\times10^{-7}$] \\
\cmidrule(lr){4-6} \cmidrule(lr){7-9}
%%%%%%%%%%%%%%%%%%%%%%%%%%%%%%%%%%%%%%%%%%%%%%%%%% 
FWHM$_{\rm [O\,\,III],1}$ & FWHM$_{\rm [O\,\,III],2}$  & $\rho$ \dotfill & 0.3 & 0.5 & 0.3 & 0.2 & 0.5 & 0.3 \\
 & & & [0.2, 0.4] & [0.4, 0.6] & [0.2, 0.4] & [0.1, 0.3] & [0.4, 0.6] & [0.2, 0.3] \\
 & & $P_{{\rm null}}$\dotfill & $2\times10^{-3}$ & $7\times10^{-8}$ & $7\times10^{-6}$ & $6\times10^{-2}$ & $2\times10^{-6}$ & $9\times10^{-4}$ \\
 & & & [$6\times10^{-5}$, $4\times10^{-2}$] & [$8\times10^{-11}$, $3\times10^{-5}$] & [$4\times10^{-8}$, $1\times10^{-3}$] & [$2\times10^{-3}$, 0.3] & [$1\times10^{-8}$, $2\times10^{-4}$] & [$6\times10^{-6}$, $2\times10^{-2}$] \\
\cmidrule(lr){4-6} \cmidrule(lr){7-9}
%%%%%%%%%%%%%%%%%%%%%%%%%%%%%%%%%%%%%%%%%%%%%%%%%%
$\frac{V_{\rm [O\,\,III],2}}{V_{\rm [O\,\,III],1}}$&$\frac{F_{\rm [O\,\,III],2}}{F_{\rm [O\,\,III],1}}$&$\rho$\dotfill & $-0.5$ & $-0.5$ & $-0.7$ & $-0.4$ & $-0.5$ & $-0.7$ \\
 & & & [$-0.5$, $-0.3$] & [$-0.6$, $-0.4$] & [$-0.8$, $-0.7$] & [$-0.5$, $-0.3$] & [$-0.6$, $-0.5$] & [$-0.7$, $-0.6$] \\
 & & $P_{{\rm null}}$\dotfill & $2\times10^{-5}$ & $1\times10^{-6}$ & $2\times10^{-30}$ & $4\times10^{-5}$ & $3\times10^{-7}$ & $3\times10^{-22}$ \\
 & & & [$1\times10^{-7}$, $2\times10^{-3}$] & [$1\times10^{-9}$, $6\times10^{-5}$] & [$4\times10^{-36}$, $7\times10^{-26}$] & [$4\times10^{-7}$, $3\times10^{-3}$] & [$4\times10^{-10}$, $2\times10^{-5}$] & [$1\times10^{-26}$, $5\times10^{-18}$]  \\
\cmidrule(lr){4-6} \cmidrule(lr){7-9}
%%%%%%%%%%%%%%%%%%%%%%%%%%%%%%%%%%%%%%%%%%%%%%%%%% 
$\frac{V_{\rm [O\,\,III],2}}{V_{\rm [O\,\,III],1}}$ & $\frac{{\rm FWHM}_{\rm [O\,\,III],2}}{{\rm FWHM}_{\rm [O\,\,III],1}}$ & $\rho$\dotfill & $0.0$ & $0.0$ & $-0.3$ & $-0.3$ & $-0.3$ & $-0.4$ \\
 & & & [$-0.2$, $0.1$] & [$-0.1$, $0.1$] & [$-0.4$, $-0.3$] & [$-0.4$, $-0.1$] & [$-0.4$, $-0.2$] & [$-0.5$, $-0.4$] \\
 & & $P_{{\rm null}}$\dotfill & 0.7 & 0.8 & $1\times10^{-5}$ & $1\times10^{-2}$ & $2\times10^{-2}$ & $6\times10^{-9}$ \\
 & & & [0.1, 0.8] & [0.2, 0.8] & [$5\times10^{-8}$, $7\times10^{-4}$] & [$7\times10^{-4}$, 0.2] & [$1\times10^{-3}$, 0.2] & [$2\times10^{-11}$, $2\times10^{-6}$]  \\
%%%%%%%%%%%%%%%%%%%%%%%%%%%%%%%%%%%%%%%%%%%%%%%%%% 
\bottomrule
\end{tabular}}
\end{center}
\end{table*}

There are at least three possible origins for the velocity splitting observed in the narrow emission lines in AGNs: galactic-scale outflows (that may or may not be associated with radio jets, e.g., \citealt{whittle88,whittle92,das06,rosario10}), mergers of stellar bulges each of which hosts a SMBH, and/or galactic disk rotation (e.g., see references in \S \ref{sec:intro}). Visual inspection of the SDSS images of the host galaxies suggests that disk rotation is likely not responsible for driving the luminosity dependence of the double-peaked fraction, since the fraction of disk components in the host galaxies of our double-peaked AGNs does not correlate with luminosity, although the AGN host galaxies are more difficult to resolve at higher redshifts. Below we focus our discussion on the possibilities that the observed luminosity dependence of the double-peaked fraction is a result of the fact that more luminous AGNs are more likely to drive galactic-scale outflows and/or to host mergers of SMBHs.

\citet{Barrows2013} have found a significant positive correlation between the velocity splittings and the quasar Eddington ratio in a sample of 131 Type 1 quasars with double-peaked \NeV\ or \NeIII\ lines selected from the SDSS at $0.8<z<1.6$; these authors have interpreted this result as evidence for radiation pressure driven outflows and that more actively accreting SMBHs drive stronger outflows.  In comparison, we have found a similar luminosity dependence in our Type 2 AGN sample in the maximum LOS splitting velocity, although the median velocity offset of our sample does not show a strong luminosity dependence (see lower left panel in Figure \ref{fig:bias}).

Figure \ref{fig:vmax} shows that the relation between $V^{{\rm max}}_{{\rm off}}$ and \OIII\ luminosity seems to be consistent with the theoretical curve as expected for radiation-pressure driven outflows with $V^{{\rm max}}_{{\rm off}}\propto L^{1/4}$ \citep[e.g.,][]{Laor2002}, except for a possible turnover at the high luminosity end (i.e., dotted line in Figure \ref{fig:vmax}). This general agreement is somewhat surprising, because the theoretical relation is expected for outflows associated with AGN broad-line regions whose sizes are at least two orders of magnitude smaller than that of the narrow-line regions being probed by the observed \OIII\ emission \citep[e.g.,][]{dietrich98,Peterson2004}. In addition, there is a significant scatter\footnote{In SDSS quasars for example, the correlation between \OIII\ luminosity and the continuum luminosity at 5100 \angstrom\ has a 0.35-dex scatter \citep{shen11}.} in the correlation between \OIII\ luminosity and the AGN bolometric luminosity \citep[e.g.,][]{heckman04,reyes08,shen11}. The apparent agreement may be: a) just a coincidence, which is unlikely, or b) resulting from the correlation between \OIII\ luminosity and host-galaxy mass due to the fact that the gas motion is confined by the depth of the host-galaxy potential well. We suggest, however, that this is unlikely to be the dominant mechanism because as shown by \citet{kauffmann03} with $z<0.3$ SDSS Type 2 AGNs that while more powerful AGNs are located preferentially in more massive host galaxies on average, the host stellar mass depends only very weakly on \loiii\ and that galaxies of given mass can host AGNs that span a very wide range (more than several orders of magnitude) in \loiii\ (see also discussion below), or c) an important piece of statistical evidence for the nature of the narrow-line velocity splitting seen in Type 2 AGNs. It may suggest that the majority of AGNs with double-peaked emission lines is indeed dominated by some galactic-scale extension of radiation-pressure driven outflows, consistent with the findings by spatially resolved followup observations at least for objects at $z\sim0.1$ \citep[e.g.,][]{Shen2011,Fu2012}.

Further evidence for AGN-driven outflows as the origin for driving the luminosity dependence observed in the double-peaked narrow line fraction comes from examinations of the dynamical properties of the emission-line gas as well as the host-galaxy stellar bulges. Figure \ref{fig:dynamics} shows that there are several correlations among the various dynamical properties in our new double-peaked sample as compared to those seen in the $z\sim0.1$ sample by \citet{Liu2010b}. Table \ref{tab:stat} lists the Spearman rank correlation coefficient $\rho$ and the null probability $P_{{\rm null}}$ for the various relations shown in Figure \ref{fig:dynamics}. Also listed are their 1$\sigma$ errors estimated from Monte Carlo simulations using Bootstrap resampling. For the sample at $z\sim0.1$ as shown by \citet{Liu2010b} and also reproduced here in Figures \ref{fig:dynamics}(a)--(d), velocity offset $V_{{\rm off}}$ is correlated with both the host-galaxy stellar velocity dispersion $\sigma_{*}$ and the FWHMs of the individual \OIII\ velocity components, and that the FWHMs are also correlated with each other. Furthermore, both the line flux and the line width ratio between the two \OIII\ velocity components are anti-correlated with the ratio of their LOS velocity offset relative to the host-galaxy systemic redshift, likely resulting from momentum conservation (Figure\ref{fig:dynamics}(e) and (f)). While our new sample at $z\sim0.5$ also shows the correlations between $V_{{\rm off}}$ and both of the two FWHMs and between the FWHMs themselves as well as the anti-correlations between the line flux (width) ratio and the LOS velocity offset ratio, there is no significant correlation between $V_{{\rm off}}$ and $\sigma_{*}$\footnote{For the new sample where robust $\sigma_{*}$ measurement is available (i.e., S/N$>5$), we have adopted the Portsmouth catalog value which was based on stellar kinematics as evaluated by the publicly available code p{\tiny PXF} \citep{Cappellari2004}. \citet{Thomas2013} have found good agreement between the Portsmouth catalog value based on p{\tiny PXF} and the ones published in SDSS DR7 which are based on the specBS pipeline \citep{SDSSDR6} as adopted by \citet{Liu2010b} for the $z\sim0.1$ double-peaked sample.} either for the whole sample of double-peaked AGNs or the subsets for which either the blueshifted velocity component is more luminous than the redshifted component (as expected for intrinsically symmetric outflows after accounting for the effect of dust extinction) or vice versa (see Table \ref{tab:stat}, Rows 1--4, for the correlation test results on the relation between $V_{{\rm off}}$ and $\sigma_{*}$ for the new sample as compared with the \citet{Liu2010b} sample). The lack of correlation between $V_{{\rm off}}$ and $\sigma_{*}$ in our new BOSS double-peaked AGN sample may be evidence that the \OIII\ gas in luminous AGNs is being highly disturbed by the AGNs (e.g., via radiation pressure driven outflows) so that the \OIII\ gas motion does not trace the potential well of the host galaxy stellar bulges \citep[e.g., see also][for a similar luminosity-dependent effect in the general population of luminous Type 2 AGNs]{Greene2009}. 

On the other hand, there is evidence that galactic-scale AGN driven outflows is likely not the sole origin for the double-peaked narrow line profiles seen in our new BOSS sample. It is possible that a small fraction of the new BOSS double-peaked AGNs contain dual AGNs, i.e., merging active SMBH pairs on $\lesssim$kpc scales. First, $\gtrsim$ half (96 out of 178) of the BOSS double-peaked sample have more prominent redshifted components, which may be more unusual under the outflow hypothesis since objects known to have biconical outflows (e.g., NGC 1068 and Mrk 78) have brighter blueshifted components \citep{axon98,heckman81,pedlar89,whittle04}. Second, as shown in Figure \ref{fig:dynamics}, the new sample exhibits similar correlations among the dynamical properties as compared with the less luminous sample at $z\sim0.1$ (Table \ref{tab:stat}) except for the lack of correlation between $V_{{\rm off}}$ and $\sigma_{*}$ as we discussed above. Spatially resolved follow up observations of the $z\sim0.1$ sample of \citet{Liu2010b} have demonstrated the mixed nature of the origins of the double-peaked narrow line profiles \citep[e.g.,][]{Shen2011,Fu2012,Comerford2015}. As an example we also shown in Figure \ref{fig:dynamics} the few cases of the confirmed cases of dual AGNs (shown as big open circles) according to \citet{Shen2011} as well as cases of NLR kinematics (shown as big open squares) involving either outflows or disk rotation. There is no clear distinction between the two populations in all of the relations being examined in Figure \ref{fig:dynamics}; the only exception is Figure \ref{fig:dynamics}(a): there is tentative evidence that the host galaxies of dual AGNs have systematically higher host-galaxy $\sigma_{*}$ at given $V_{{\rm off}}$ than the alternative cases of NLR kinematics, although the sample statistics is still too poor to draw any firm conclusion. Nevertheless, Figure \ref{fig:dynamics} demonstrates that the new BOSS sample spans the whole range of parameter space covered by the $z\sim0.1$ sample in terms of the various emission-line dynamical properties. Therefore similar to the $z\sim0.1$ sample, it is likely that the new BOSS sample also contains cases of both AGN driven outflows, dual AGNs, and/or disk rotation. Finally, as already noted by several studies \citep[e.g.,][]{Liu2010a,Shen2011,Fu2012,Comerford2015}, it is possible to have multiple cases in one galaxy since they are not mutually exclusive (e.g., dual AGN systems may have disk rotation contributing to the narrow-line velocity splitting, \citealt{barnes02,Blecha2013a}, and/or to AGN/merger driven outflows which are not uncommon in galaxy mergers).

Spatially resolved follow up observations are needed to resolve the origin(s) for the double-peaked narrow line profile seen in the new BOSS sample for individual AGNs. The results would be of interest to galaxy formation and evolution for understanding the significance of possible SMBH feedback \citep[e.g.,][see \citealt{Fabian2015} for a review]{greene11,Liugl2013,Liugl2013b,Greene2012,Greene2014,Crenshaw2015} and the effects of galaxy mergers in triggering or enhancing AGNs \citep[e.g.,][]{koss10,ellison11,Liu2012,koss12,teng12,vanwassenhove12,Steinborn2015}. They would also help constraining the abundance of low-frequency gravitational wave sources from merging SMBHs following galaxy mergers \citep[e.g.,][]{haehnelt94,yu02,komossa03b,volonteri03,Volonteri2009,Dotti2012,DEGN,Colpi2014,Kelley2016} relevant for current pulsar timing arrays \citep[e.g.,][]{hobbs10,Sesana2012,Rosado2013,Burke-Spolaor2013,Babak2015,Huerta2015,Shannon2015,Middleton2016} and future space-based experiments \citep[e.g.,][]{holz05,cornish07,hughes09,Centrella2010,Babak2011,amaro12,Klein2016,Tamanini2016}. In particular, the dominant uncertainty in the abundance of low-frequency gravitational wave signals from merging SMBHs is related to our poor understanding the astrophysical links between merging SMBHs and their host galaxies and to the physical processes that drive SMBH binaries to the gravitational wave dominated regime \citep[e.g.,][]{Roebber2016,Komossa2015a}. 

Statistical studies of double-peaked narrow lines in AGN have mostly been focusing on emission-line galaxies at $z\sim0.1$ (e.g., references in \S \ref{sec:intro}). More work is still needed at higher redshifts significantly larger than 0.1 \citep[e.g.,][]{Takada2014} where outflows and dual AGNs are likely to be more common. The new sample presented in this work represents the first statistically robust constraint on the fraction of AGNs with narrow-line double peaks at intermediate redshift based on \OIII\ selection. \citet{Comerford2013} have searched for double-peaked narrow-line AGNs in the AGN and Galaxy Evolution Survey \citep{Kochanek2012} and have found two double-peaked objects among 173 Type 2 AGNs at $z<0.37$. \citet{Yuan2016} have also presented a sample of 2758 Type 2 quasars at $z<1$ from the SDSS-III/BOSS that include 654 candidate double-peaked \OIII\ profiles. While our parent Type 2 AGN sample has been selected based on AGN diagnostic emission-line ratios at all redshift being considered (\S \ref{subsubsec:agn} and Figure \ref{fig:bpt}), the \citet{Yuan2016} sample was selected based on diagnostic emission-line ratios at $z<0.52$ and detection of \NeV\ at $z>0.52$. Unlike \citet{Yuan2016} which focus on Type 2 quasars, i.e., the most luminous population among all Type 2 AGNs, our parent AGN sample selection does not contain any minimum threshold cut on \OIII\ EW or luminosity, although our criteria for the selection of double-peaked \OIII\ profiles is likely more stringent than that of \citet{Yuan2016} and we generally do not include more complex profiles such as lumpy, winged, or multi-component velocity signatures (\S \ref{subsubsec:dp}). 

Besides studies on Type 2 AGNs/quasars, \citet{Barrows2013} have presented a sample of 131 quasars (i.e., with broad emission lines or Type 1 objects) from the SDSS at $0.8<z<1.6$ with double peaks in either of the high-ionization narrow emission lines \NeV\ or \NeIII , representing the first attempt at extending the selection of AGNs with double-peaked narrow emission lines to higher redshifts. While the selection based on \NeV\ or \NeIII\ lines allows for the identification of AGNs at higher redshifts, the \citet{Barrows2013} sample, however, is highly incomplete due to the intrinsic weakness of the emission lines. Nevertheless, their result on the fraction of Type 1 quasars with double-peaked narrow emission lines generally agrees with our result for the most luminous Type 2 AGNs after correcting for luminosity-dependent selection incompleteness (which is only about 1\% complete at emission-line EW=10, \citealt{Barrows2013}), although the authors focused on the redshift evolution of the double-peaked fraction without discussing the luminosity dependence. This overall agreement in the fraction of Type 1 and Type 2 AGNs that exhibit double-peaked narrow emission lines is perhaps not surprising considering AGN unification \citep{antonucci93,urry95} and in particular in the quasar luminosity regime \citep[e.g.,][]{zakamska04,zakamska06,reyes08}. The combination of the \citet{Liu2010b} sample with the new sample presented in this work, which is also based on \OIII , allows us to make a more robust determination of the luminosity dependence of the double-peaked fraction less affected by systematic uncertainties induced by selection based on different emission lines. Only with the statistical power of the SDSS-III/BOSS survey are we able to cover enough volume to identify these luminous and rare objects and address the luminosity dependence of their occurrence rate in a statistically robust sense.

%
%------------------------------------------------------------------------------
%\acknowledgments
\section*{Acknowledgements}

We thank Yue Shen for inspiring us to mine the SDSS-III database, Laura Blecha and Yue Shen for helpful discussion, and an anonymous referee for a
prompt and careful report that improves the paper.

We are grateful to the SDSS-III Portsmouth group for making their spectral measurements publicly available.

Funding for SDSS-III has been provided by the Alfred P. Sloan Foundation, the Participating Institutions, the National Science Foundation, and the U.S. Department of Energy Office of Science. The SDSS-III web site is http://www.sdss3.org/.

SDSS-III is managed by the Astrophysical Research Consortium for the Participating Institutions of the SDSS-III Collaboration including the University of Arizona, the Brazilian Participation Group, Brookhaven National Laboratory, Carnegie Mellon University, University of Florida, the French Participation Group, the German Participation Group, Harvard University, the Instituto de Astrofisica de Canarias, the Michigan State/Notre Dame/JINA Participation Group, Johns Hopkins University, Lawrence Berkeley National Laboratory, Max Planck Institute for Astrophysics, Max Planck Institute for Extraterrestrial Physics, New Mexico State University, New York University, Ohio State University, Pennsylvania State University, University of Portsmouth, Princeton University, the Spanish Participation Group, University of Tokyo, University of Utah, Vanderbilt University, University of Virginia, University of Washington, and Yale University.

Facilities: Sloan

%%%%%%%%%%%%%%%%%%%%%%%%%%%%%%%%%%%%%%%%%%%%%%%%%%



%\clearpage

\bibliographystyle{/Users/zeus/Documents/References/mnras}
\bibliography{/Users/zeus/Documents/References/binaryrefs}

%%%%%%%%%%%%%%%%%%%%%%%%%%%%%%%%%%%%%%%%%%%%%%%%%%

%%%%%%%%%%%%%%%%% APPENDICES %%%%%%%%%%%%%%%%%%%%%

%\appendix

%%%%%%%%%%%%%%%%%%%%%%%%%%%%%%%%%%%%%%%%%%%%%%%%%%

% Don't change these lines
\bsp	% typesetting comment
\label{lastpage}
\end{document}